\documentclass[pre,twocolumn]{revtex4}

\usepackage{graphicx}
\usepackage{dcolumn}
\usepackage{amsmath}
\usepackage{amssymb}
\usepackage{amsfonts}

\newcommand{\dotprod}{{\scriptscriptstyle \stackrel{\bullet}{{}}}}

\begin{document}

\title{Pattern Formation and Dynamics in Rayleigh-B\'{e}nard Convection:
Numerical Simulations of Experimentally Realistic Geometries}

\author{M.R. Paul}
\email{mpaul@caltech.edu}
  \homepage{http://www.cmp.caltech.edu/~stchaos}
\author{K.-H. Chiam}
\author{M.C. Cross}
\affiliation{Department of Physics, California Institute of
Technology 114-36, Pasadena, California 91125}

\author{P.F. Fischer}
\affiliation{Mathematics and Computer Science Division, Argonne
National Laboratory, Argonne, Illinois 60439}

\author{H. S. Greenside}
\affiliation{Department of Physics, Duke University, Durham, North
Carolina 27708-0305}

\date{\today}

\begin{abstract}
Rayleigh-B\'{e}nard convection is studied and quantitative
comparisons are made, where possible, between theory and
experiment by performing numerical simulations of the Boussinesq
equations for a variety of experimentally realistic situations.
Rectangular and cylindrical geometries of varying aspect ratios
for experimental boundary conditions, including fins and spatial
ramps in plate separation, are examined with particular attention
paid to the role of the mean flow. A small cylindrical convection
layer bounded laterally either by a rigid wall, fin, or a ramp is
investigated and our results suggest that the mean flow plays an
important role in the observed wavenumber. Analytical results are
developed quantifying the mean flow sources, generated by
amplitude gradients, and its effect on the pattern wavenumber for
a large-aspect-ratio cylinder with a ramped boundary. Numerical
results are found to agree well with these analytical predictions.
We gain further insight into the role of mean flow in pattern
dynamics by employing a novel method of quenching the mean flow
numerically. Simulations of a spiral defect chaos state where the
mean flow is suddenly quenched is found to remove the time
dependence, increase the wavenumber and make the pattern more
angular in nature.
\end{abstract}


\maketitle

\section{Introduction}
\label{section:introduction} Rayleigh-B\'{e}nard convection has
played a crucial role in guiding both theory and experiment
towards an understanding of the emergence of complex dynamics from
nonequilibrium systems~\cite{cross:1993}. However, an important
missing link has been the ability to make quantitative and
reliable comparisons between theory and experiment.

Nearly all previous three-dimensional convection calculations have
been subject to a variety of limitations.  Many simulations have
been for small aspect ratios where the lateral boundaries dominate
the dynamics, and as a result, complicate the analysis. When
larger aspect ratios are considered, it is often with the
assumption of periodic boundaries, which is convenient numerically
yet does not correspond to any laboratory experiment. As a result
of algorithmic inefficiencies, or the lack of computer resources,
simulations have frequently not been carried out for long times.
This presents the difficulty in determining whether the observed
behavior represents the asymptotic non-transient state, which is
usually the state that is most easily understood theoretically.

Fortunately, advances in parallel computers, numerical algorithms
and data storage are such that direct numerical simulations of the
full three-dimensional time dependent equations are possible for
experimentally realistic situations. We have performed simulations
with experimentally correct boundary conditions, in geometries of
varying shapes and aspect ratios over long enough times so as to
allow a detailed quantitative comparison between theory and
experiment.

Alan Newell has made numerous important contributions to the
discussion of pattern formation in non-equilibrium systems. In
this paper, presented in this special issue in his honor, we give
a survey of our recent results that touch on many of the issues he
has raised, and in turn make use of some of the tools that he has
helped develop to understand our simulations.

\section{Simulation of Realistic Geometries}
\label{section:numerical simulation} We have performed full
numerical simulations of the fluid and heat equations using a
parallel spectral element algorithm (described in detail elsewhere
\cite{fischer:1997}). The velocity $\vec{u}$, temperature $T$, and
pressure $p$, evolve according to the Boussinesq equations,
\begin{eqnarray}
  {\sigma}^{-1} \left(
  {\partial}_t + \vec{u} \dotprod \vec{\nabla} \right) \vec{u}
  &=&  -\vec{\nabla} p + RT \hat{z} + \nabla^2 \vec{u}  , \label{eq:mom}\\
  \left( {\partial}_t + \vec{u} \dotprod \vec{\nabla} \right) T
  &=& \nabla^2 T  , \label{eq:energy}\\
  \vec{\nabla} \dotprod \vec{u} &=& 0,
  \label{eq:mass}
\end{eqnarray}
where $\partial_t$ indicates time differentiation, $\hat{z}$ is a
unit vector in the vertical direction opposite of gravity, $R$ is
the Rayleigh number, and $\sigma$ is the Prandtl number. The
equations are nondimensionalized in the standard manner using the
layer height $h$, the vertical diffusion time for heat ${\tau}_v
\equiv h^2/\kappa$ where $\kappa$ is the thermal diffusivity, and
the temperature difference across the layer $\Delta T$, as the
length, time, and temperature scales, respectively.

We have investigated a wide range of geometries including
cylindrical and rectangular domains, which are the most common
experimentally, in addition to elliptical and annular domains.
Rotation about the vertical axis of the convection layer for any
of these situations is also possible but will not be presented
here. All bounding surfaces are no-slip, $\vec{u}=0$, and the
lower and upper surfaces and are held at constant temperature,
$T(z=0)=1$ and $T(z=1)=0$.

\begin{figure}[tbh]
\begin{center}
\includegraphics[width=2.5in]{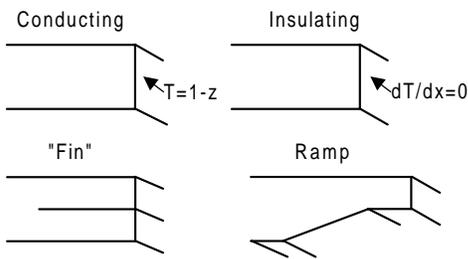}
\end{center}
\caption{Four lateral sidewall boundary conditions utilized in the
numerical simulations. The two thermal boundary conditions are
conducting and insulating whereas the fin and ramp represent
geometric conditions employed in experiments.}
\label{fig:sidewalls}
\end{figure}

A variety of sidewall boundary conditions are shown in
Fig.~\ref{fig:sidewalls}. Common thermal boundary conditions on
the lateral sidewalls are insulating, $\hat{n} \dotprod
\vec{\nabla}T = 0$ where $\hat{n}$ is a unit vector normal to the
boundary at a given point, and conducting, $T=1-z$. In the future
we will have the flexibility of imposing a more experimentally
accurate thermal boundary condition by coupling the fluid to a
lateral wall of finite thickness and known finite thermal
conductivity that is bounded on the outside by a vacuum.

In experiment, however, small sidewall thermal forcing can have a
significant effect upon the resulting patterns and, as a result,
finned boundaries have been
employed~\cite{daviaud:1989,debruyn:1996,pocheau:1997}. These are
formed by inserting a very thin piece of paper or cardboard
between the top and bottom plates near the sidewalls. This
suppresses convection over the finned region ($R \sim h^3$ and the
layer height has effectively been reduced) whereas in the bulk of
the domain, i.e. the un-finned region, supercritical conditions
prevail. This is accomplished numerically by extending a no-slip
surface into the domain from the lateral sidewall. In all of our
simulations we have chosen the vertical position of the fin to be
$z=0.5$ but this is not necessary. The result is that the
supercritical portion of the convection layer is bounded by a
subcritical region of the same fluid and hence with the same
material properties. An additional effect is that the mean flow
may extend into the finned region which presents an interesting
scenario for exploring the effect of mean flows upon pattern
dynamics~that has been investigated both experimentally and
theoretically by Pocheau and Daviaud~
\cite{daviaud:1989,{pocheau:1997}} and is discussed further below.

The sidewalls can also have an orienting effect and ramped
boundaries have been used as a ``soft boundary"~\cite{kramer:1982}
in an effort to minimize this. By gradually decreasing the plate
separation as the lateral sidewall is approached the convection
layer eventually becomes critical and then increasingly
subcritical. Using the spectral element algorithm we are able to
investigate arbitrary ramp shapes: we have chosen to investigate
the precise radial ramp utilized in recent
experiments~\cite{bajaj:1999,{ahlers:2001}} on a cylindrical
convection layer. Again the mean flow is able to extend into the
subcritical region.

\begin{figure}[tbh]
\begin{center}
\includegraphics[width=2.5in]{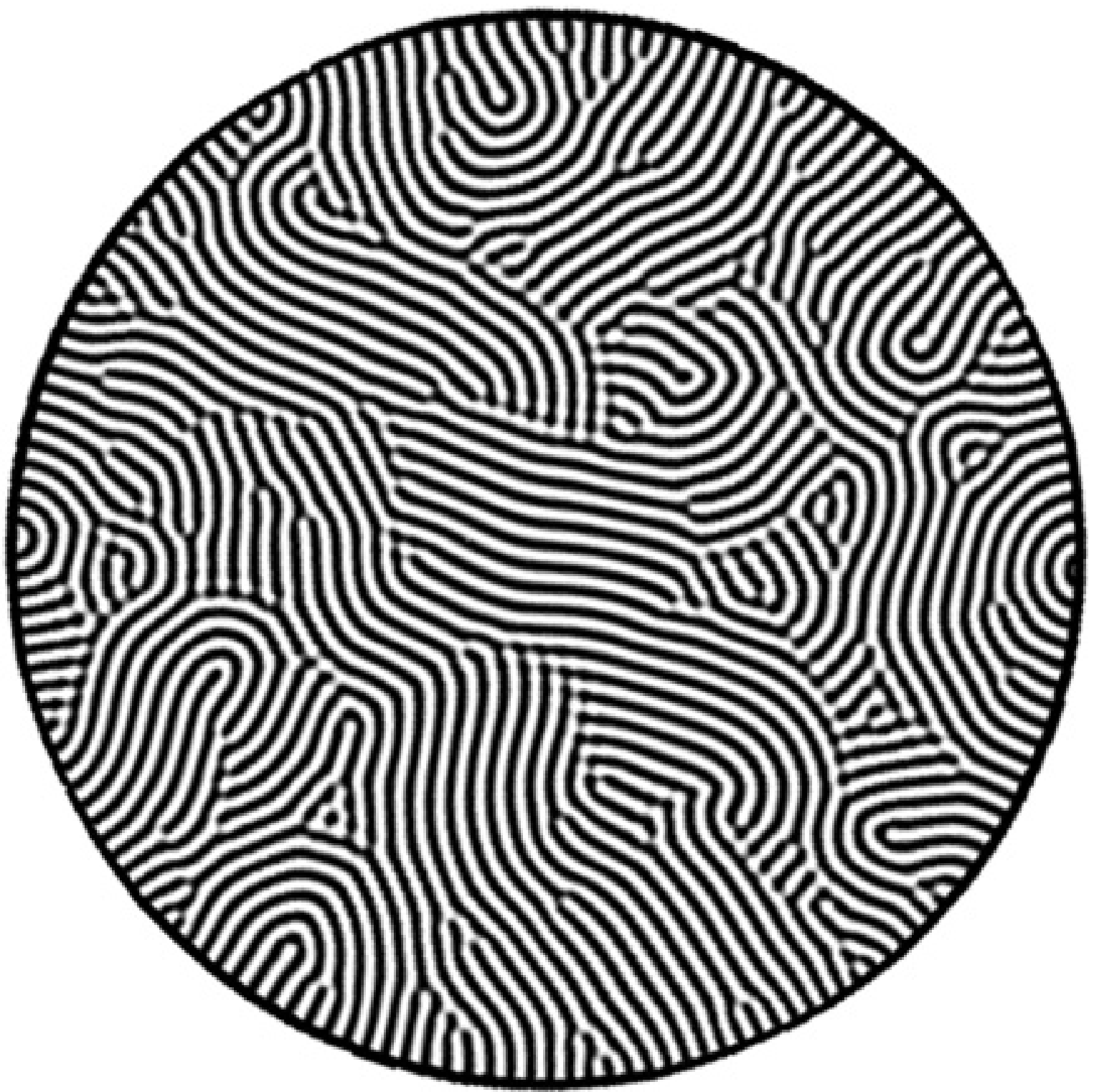}
\includegraphics[width=1.32in]{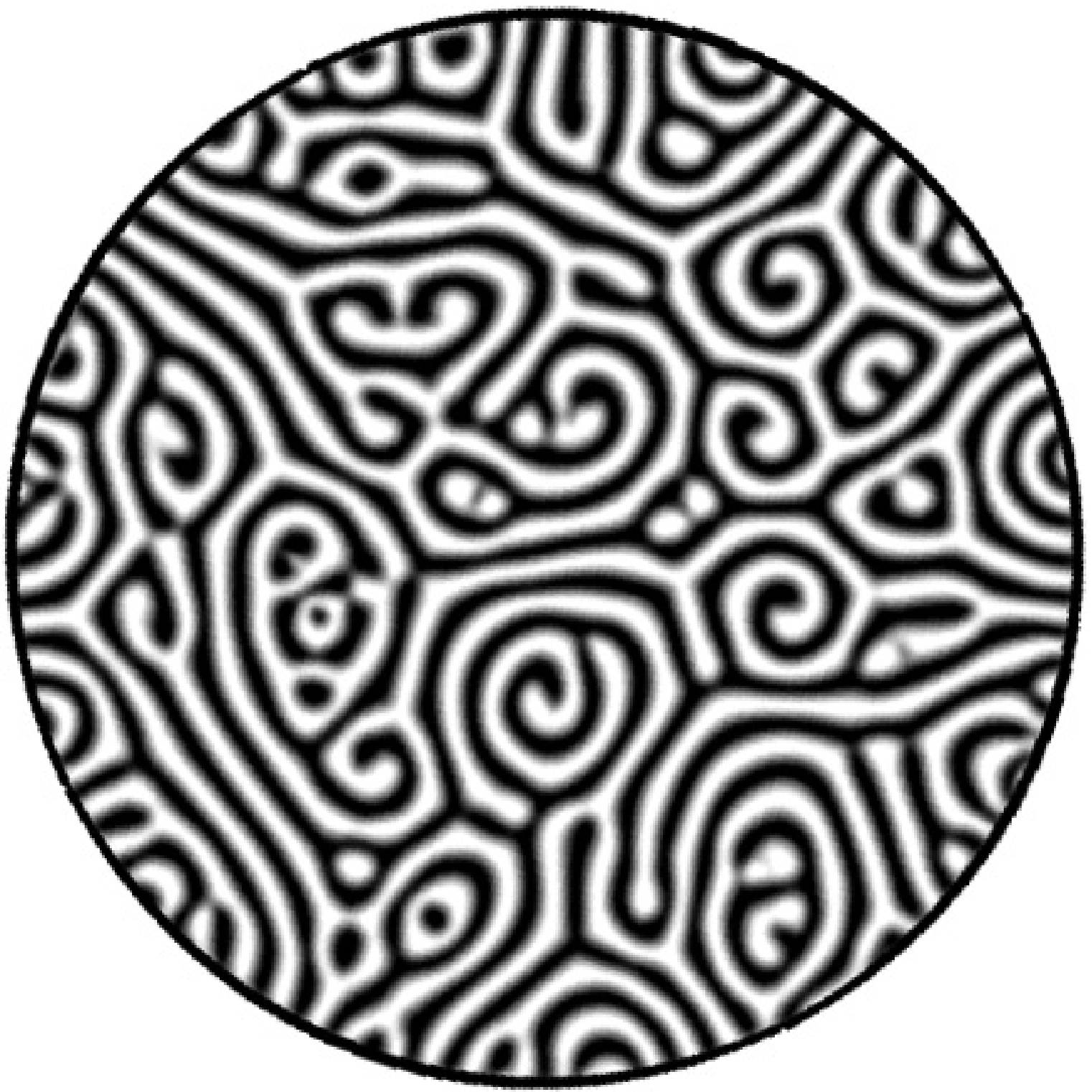}
\end{center}
\caption{Numerical simulations of two large-aspect-ratio
cylindrical convection layers. The pattern is illustrated by
contours of the thermal perturbation, dark regions represent cool
descending fluid and light regions warm ascending fluid. Both
simulations are initiated from random thermal perturbations and
the lateral sidewalls are insulating. (Top) $\Gamma=57$,
$\sigma=2.94$, $R=2169.2$ and $t=74 \tau_v$. (Bottom) A spiral
defect chaos state, $\Gamma=30$, $\sigma=1.0$, $R=2950$ and $t=254
\tau_v$. } \label{fig:large_gamma}
\end{figure}

Perhaps the most common method employed experimentally to reduce
the influence of sidewalls is to use a large aspect ratio
$\Gamma$, where $\Gamma=r/h$ in a cylindrical domain where $r$ is
the radius and $\Gamma=L/h$ in a square domain where $L$ is the
length of side. Experiments can attain aspect ratios as large as
$\sim 500$. However, the majority of large aspect ratio
experiments are for $\Gamma \lesssim 100$. We have performed
numerical simulations using the spectral element algorithm for
$\Gamma \sim 60$ as shown in Fig.~\ref{fig:large_gamma}.

The top panel in Fig.~\ref{fig:large_gamma} illustrates the
convection pattern present for the parameters of the classic
paper~\cite{ahlers:1974} where flow visualization was not
possible. Although the simulation has only been performed for a
short time $t_f \sim 100 \tau_v$ it appears that a slow process of
domain coarsening~\cite{cross:1995:physrevlett} is occurring. The
bottom of Fig.~\ref{fig:large_gamma} illustrates the time
dependent spatiotemporal chaotic state of spiral defect
chaos~\cite{morris:1993}. These, and other, interesting large
aspect ratio problems can now be addressed through the use of
numerical simulation.

Heuristically, using the spectral element algorithm on an IBM SP
parallel supercomputer, it is our experience that it is practical
to perform full numerical simulations for aspect ratios $\Gamma
\sim 30$ for simulation times of $t_f \sim \tau_h$ (36 hours on 64
processors), where $\tau_h$ is the horizontal diffusion time for
heat ${\tau}_h={\Gamma}^2 {\tau}_v$, and $\Gamma \sim 60$ for $t_f
\sim 300 \tau_v$ (36 hours on 256 processors) for $\epsilon
\lesssim 1$, $0.5 \lesssim \sigma \lesssim 10$, $\Delta t \approx
0.01$, and approximately cubic shaped spectral elements with an
edge length of unity and $11^{th}$ order polynomial expansions
(where $\epsilon=(R-R_c)/R_c$ and $R_c$ is the critical value of
the Rayleigh number). Of course for smaller domains the
computational requirements significantly decrease.

A major benefit of numerical simulations is that a complete
knowledge of the flow field is produced. For example, we have
first used this to address a long standing open question
concerning chaos in small cylindrical domains. The existence of a
power-law behavior in the fall-off of the power spectral density
derived from a time series of the Nusselt number was not
understood~\cite{ahlers:1974}. The Nusselt number, $N(t)$, is a
global measurement of the temperature difference across the fluid
layer. In cryogenic experiments very precise measurements of
$N(t)$ are
possible~\cite{ahlers:1974,{ahlers:1978},{ahlers:1980},{libchaber:1978}},
however the flow field can not be visualized easily. Subsequent
room temperature experiments using compressed gasses allowed flow
visualization at the expense of precise measurements of the
Nusselt
number~\cite{pocheau:1985,{croquette:1986},{pocheau:1989}}.

\begin{figure}[tbh]
\begin{center}
\includegraphics[width=2.5in]{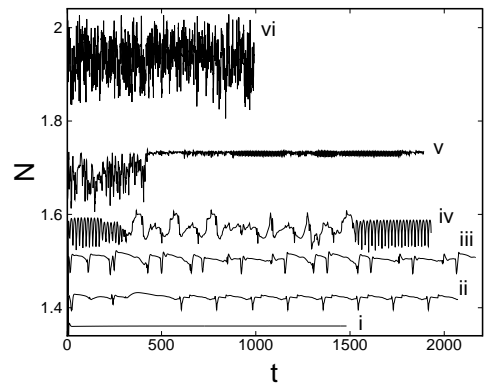}
\end{center}
\caption{Plots of the dimensionless heat transport N(t) for
reduced Rayleigh number $\epsilon = 0.557,0.614,0.8,1.0,1.5$, and
$3.0$, labelled (i-vi) respectively ($\Gamma = 4.72$). For cases
(i-v), $\Delta t =0.01$, and for case~(vi), $\Delta t = 0.005$
($\Delta t$ is the time step).} \label{fig:nu_all_cond}
\end{figure}

By performing long-time simulations, on the order of many
horizontal diffusion times, for the same parameters in cylindrical
domains with $\sigma=0.78$ and for a range of $\epsilon$, with
realistic boundary conditions, we had access to both precise
measurements of the Nusselt number, Fig.~\ref{fig:nu_all_cond},
and flow visualization, Fig.~\ref{fig:snapshots_mf}, allowing us
to resolve the issue~\cite{paul:2001}. Conducting sidewalls were
used and all simulations were initiated from small, $\delta T
\approx 0.01$, random thermal perturbations. Flow visualization of
the simulations represented in Fig.~\ref{fig:nu_all_cond} display
a rich variety of dynamics similar to what was observed in the
room temperature experiments. Using simulation results, the
particular dynamical events responsible for the $N(t)$ signature
were identified. The power-law behavior was found to be caused by
the nucleation of dislocation pairs and roll pinch-off events.
Additionally, the power spectral density was found to decay
exponentially for large frequencies as expected for
time-continuous deterministic dynamics. The large frequency regime
was not accessible to experiment because of the presence of the
noise floor.
\section{Role of mean flow} \label{section:mean_flow} The mean flow
present in these flow fields, and in general for $\sigma \lesssim
1$, plays an important role in
theory~\cite{newell:1990:jfm,{cross:1984}} yet it is not possible
to measure or visualize the mean flows in the current generation
of experiments. In our simulations, however, we can quantify and
visualize the mean flow.
\begin{figure}[tbh]
\begin{center}
\includegraphics[width=2.5in]{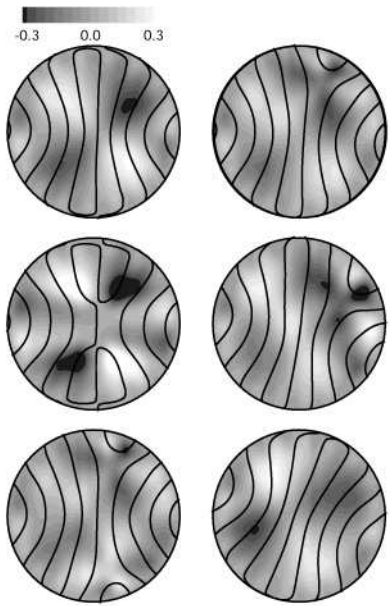}
\end{center}
\caption{Flow visualization showing the pattern (solid dark lines)
and shaded contours of the vorticity potential, $\zeta$, for
$\epsilon=0.614$ labelled ii) in Fig~\ref{fig:nu_all_cond}
($\Gamma = 4.72$). Dark regions corresponding to negative
vorticity generate clockwise mean flow and light regions to a
positive vorticity generating a counter clockwise mean flow. The
dark solid lines are zeros of the thermal perturbation at
mid-depth illustrating the outline of the convection rolls. From
top to bottom and left to right the panels are for $t =
600,605,630,650,735,785$. The dislocations glide toward the right
wall focus (shown here); during the next half period, the
dislocations glide to the left wall focus. This left and right
alternation continues for the entire simulation.}
\label{fig:snapshots_mf}
\end{figure}

The mean flow field, $\vec{U}(x,y)$, is the horizontal velocity
integrated over the depth and originates from the Reynolds stress
induced by pattern distortions. Recalling the fluid equations,
Eqs.~(\ref{eq:mom}) and~(\ref{eq:mass}), it is evident that the
pressure is not an independent dynamic variable. The pressure is
determined implicitly to enforce incompressibility,
\begin{equation} \label{eq:pressure}
\nabla^2 p = -\sigma^{-1} \vec{\nabla} \dotprod \left[ \left(
\vec{u} \dotprod \vec{\nabla} \right) \vec{u} \right] + R
\partial_z T.
\end{equation}
Focussing on the nonlinear Reynolds stress term and rewriting the
pressure as $p = p_o(x,y) + \bar{p}(x,y,z)$ yields,
\begin{equation} \label{eq:pressure_p0}
p_o(x,y) \sim \sigma^{-1} \int dx' dy' \ln \left( 1 / \left| r-r'
\right| \right) \left< \vec{\nabla}' \dotprod \left[ \left(
\vec{u} \dotprod \vec{\nabla} \right) \vec{u} \right] \right>_z .
\end{equation}
In Eq.~(\ref{eq:pressure_p0}) the $\ln(1/|r-r'|)$ is not exact, in
order to be more precise the finite system Green's function would
be required. However, the long range behavior persists. This gives
a contribution to the pressure that depends on distant parts of
the convection pattern. The Poiseuille-like flow driven by this
pressure field subtracts from the Reynolds stress induced flow
leading to a divergence free horizontal flow that can be described
in terms of a vertical vorticity.

The mean flow is important not because of its strength; under most
conditions the mean flow is substantially smaller than the
magnitude of the roll flow making it extremely difficult to
quantify experimentally. The mean flow is important because it is
a nonlocal effect acting over large distances (many roll widths)
and changes important general predictions of the phase
equation~\cite{cross:1984}. The mean flow is driven by roll
curvature, roll compression and gradients in the convection
amplitude. The resulting mean flow advects the pattern, giving an
additional slow time dependence.

The mean flow present in the simulation flow fields,
$\vec{U}_s(x,y)$, is formed by calculating the depth averaged
horizontal velocity,
\begin{equation}
\vec{U}_s(x,y)=\int^1_0 \vec{u}_{\perp}(x,y,z)dz
\label{eq:mean_flow_sim}
\end{equation}
where $\vec{u}_{\perp}$ is the horizontal velocity field.
Furthermore it will be convenient to work with the vorticity
potential, $\zeta$, defined as
\begin{equation} \label{eq:vorticity_potential}
\nabla^2_\perp \zeta = -\hat{z} \dotprod \left(
\vec{\nabla}_{\perp} \times \vec{U}_s \right) = - \omega_z
\end{equation}
where $\omega_z$ is the vertical vorticity and $\nabla^2_\perp$ is
the horizontal Laplacian.

Six consecutive snapshots in time for the periodic dynamics shown
in Fig.~\ref{fig:nu_all_cond} case ii) are illustrated in
Fig.~\ref{fig:snapshots_mf}. One half period is displayed
illustrating the nucleation of a dislocation pair and its
subsequent annihilation in the opposing wall foci. The vorticity
potential, $\zeta$, is shown on a grey scale: dark regions
represent negative vorticity and light regions represent positive
vorticity which will generate a clockwise and a counter clockwise
rotating mean flow, respectively. The quadrupole spatial structure
of $\zeta$ in the first panel, i.e. four lobes of alternating
positive and negative vorticity with one lobe per quadrant,
generates a roll compressing mean flow that pushes the system
closer to a dislocation pair nucleation event. During dislocation
climb and glide the spatial structure of the vorticity potential
is more complicated until the pan-am pattern is reestablished in
final panel and a quadrupole structure of vorticity is again
formed and the process repeats. The dislocations alternate gliding
left and right resulting is a slight rocking back and forth of the
entire pattern with each half period which is visible in the
different pattern orientations in the first last panels. This
alternation persists for the entire simulation.

A numerical investigation of the importance of the mean flow for
this small cylindrical domain was performed by implementing the
ramped and finned boundary conditions. In all of these simulations
the bulk region of constant $R$ extended out to a radius
$r_0=4.72$. In the finned case a fin at half height occupied the
region $4.72 \le r \le 7.66$. In the ramped case a radial ramp in
plate separation was given by,
\begin{equation}
 h(r) = \left\{ \begin{array}{ll}
   1, & \mbox{$r < r_0$} \\
   1 - {\delta_r} \left[ 1- \cos \left( \frac{r-r_0} {r_1-r_0} \pi \right) \right], & \mbox{$r \ge
   r_0$}
\end{array}\right.
\label{eq:platesep}
\end{equation}
where $r_0=4.72$, $r_1=10.0$, and $\delta=0.15$.

\begin{figure}[tbh]
\begin{center}
\includegraphics[width=3.0in]{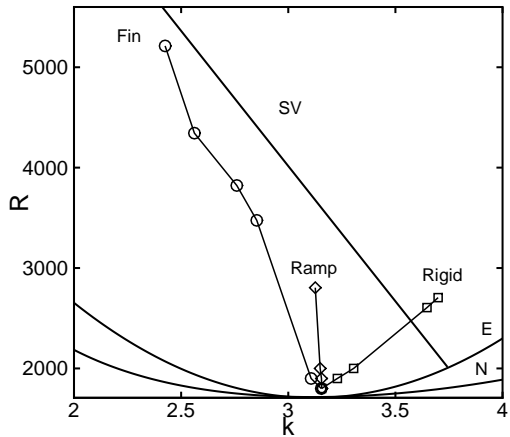}
\end{center}
\caption{Mean pattern wavenumber measurements for a cylindrical
convection layer, $r_0=4.72$, with rigid sidewalls ($\Box$), fin
($\circ, r_1=7.66$) and a spatial ramp in plate separation
($\diamond, r_1=10.0$, $r_c=7.34$, and  $\delta=0.15$). In all
three cases the sidewalls are perfectly conducting and
$\sigma=0.78$. For reference, solid lines labelled E, N, and SV
indicate the approximate location of the Eckhaus, Neutral and
Skewed Varicose stability boundaries for an infinite layer
straight parallel convection rolls. All patterns represented are
time independent.} \label{fig:wnall_smallgamma}
\end{figure}

\begin{figure}[tbh]
\begin{center}
\includegraphics[width=2.0in]{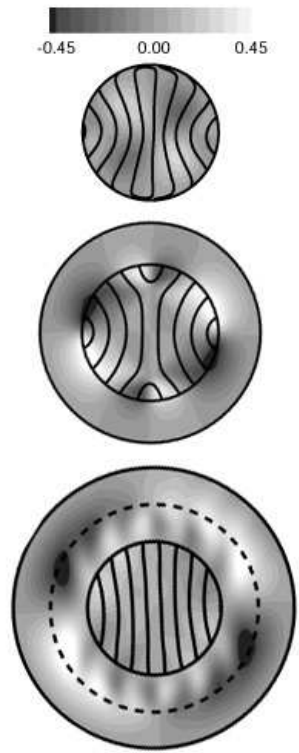}
\end{center}
\caption{Convection pattern and shaded contours of the vorticity
potential, $\zeta$, for a cylindrical convection layer,
$r_0=4.72$, with rigid sidewalls, fin ($r_1=7.66$) and a spatial
ramp in plate separation ($r_1=10.0$, $r_c=7.34$ and marked with a
dashed line, and $\delta=0.15$) shown top, middle and bottom,
respectively. The convection pattern is illustrated by plotting
zero contours of the thermal perturbation. In all three cases the
sidewalls are perfectly conducting, $\sigma=0.78$ and $R=2804$.}
\label{fig:rigid_fin_ramp}
\end{figure}

The different mean wavenumber behavior (using the Fourier methods
discussed in \cite{morris:1993}) exhibited in these three
different cases is shown in Fig.~\ref{fig:wnall_smallgamma}. As
illustrated in Fig.~\ref{fig:rigid_fin_ramp} the behavior of the
vorticity potential suggests an explanation. In the simulations
with a rigid sidewall, not ramped or finned, the vorticity
potential generates a mean flow that enhances roll compression, as
described above. In the case of the finned and ramped boundaries
the vorticity potential and the resulting mean flow are being
generated by gradients in the convection amplitude and are largely
situated away from the bulk of the domain. Furthermore, the mean
flow generated is strongest in the subcritical finned or ramped
region away from the convection rolls. This is demonstrated by
comparing the average value of the mean flow over a fraction of
the bulk of the domain, $r \le 1$, where it was found that
$\bar{U}_s = $ 0.23, 0.09, and 0.02 for the rigid, finned and
ramped domains, respectively, and that the maximum flow field
velocity is $|\vec{u}| \approx 10$.

It is attractive to pursue the case of a radial ramp in plate
separation because the variation in the convective amplitude
caused by the ramp can be determined analytically and the
influence of a mean flow upon nearly straight rolls can be
quantified~\cite{paul:2002:pre}. Usually the mean flow can only be
determined once the texture is known and it is hard to calculate
because of defects acting as sources, in addition to the regions
of smooth distortions.

\begin{figure}[tbh]
\begin{center}
\includegraphics[width=3.5in]{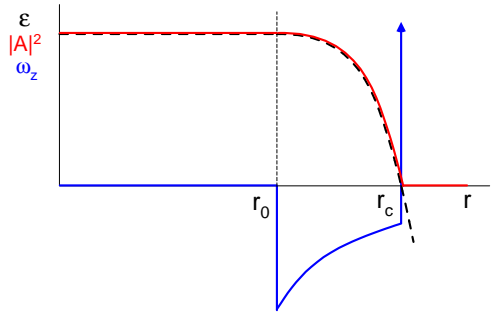}
\end{center}
\caption{A schematic illustrating the radial variation, for purely
adiabatic conditions, of $\epsilon$ (dashed line), $|A|^2$ (solid
line), and $\omega_z$ (solid line with arrow) for a cylindrical
convection layer with a radial ramp in plate separation. Labelled
$r_0$ and $r_c$ are the radial values where the ramp begins and
where the ramp yields critical conditions, respectively. Note that
for $r>r_c$, $|A|^2=0$ in the adiabatic approximation.}
\label{fig:ramp_vorticity_schematic}
\end{figure}

Near threshold an explicit expression for the mean flow,
$\vec{U}$, that advects the convection pattern
is~\cite{cross:1984}
\begin{equation}
\vec{U}(x,y) = - \gamma \vec{k} \vec{\nabla}_{\perp} \dotprod
\left( \vec{k} |A|^2 \right) - \vec{\nabla}_{\perp}p_o(x,y)
\label{eq:meanflow}
\end{equation}
where $\gamma$ is a coupling constant given by $\gamma = 0.42
{\sigma}^{-1} (\sigma+0.34)(\sigma+0.51)^{-1}$, $|A|^2$ is the
convection amplitude normalized so that the convective heat flow
per unit area relative to the conducted heat flow at $R_c$ is
$|A|^2R/R_c$, $p_o$ is the slowly varying pressure (see
Eq.~(\ref{eq:pressure_p0})) and $\vec{\nabla}_{\perp}$ is the
horizontal gradient operator. The vertical vorticity is then given
by the vertical component of the curl of Eq.~(\ref{eq:meanflow}),
\begin{equation}
{\omega}_z = \hat{z} \dotprod \left( \vec{\nabla}_{\perp} \times
\vec{U} \right) = - \gamma \hat{z} \dotprod \vec{\nabla}_{\perp}
\times \left[ \vec{k} \vec{\nabla}_{\perp} \dotprod \left( \vec{k}
|A|^2 \right) \right]. \label{eq:wz_general}
\end{equation}
Consider a cylindrical convection layer with a radial ramp in
plate separation containing a field of x-rolls given by
$\vec{k}=k_o \hat{x}$. The amplitude can be represented for large
$\epsilon_o$, using an adiabatic approximation, as
$|A|^2=\epsilon(r)/g_o$ for $\epsilon>0$ and $|A|^2=0$ for
$\epsilon(r)<0$ as shown in
Fig.~\ref{fig:ramp_vorticity_schematic}, making the amplitude a
function of radius only $|A|^2=f(r)$.
Inserting $|A|^2=f(r)$ into Eq.~(\ref{eq:wz_general}) yields,
after some manipulation, the following expression for the vertical
vorticity,
\begin{equation}
{\omega}_z = \frac{\gamma {k_o}^2}{2} \left[ \frac{d^2|A|^2}{dr^2}
- \frac{1}{r} \frac{d|A|^2}{dr} \right] \sin 2 \theta.
\label{eq:wz}
\end{equation}
The vorticity generated by the amplitude variation caused by the
ramp is also shown in Fig.~\ref{fig:ramp_vorticity_schematic}:
there is a negative vorticity for $r_0<r<r_c$ and then a delta
function spike of positive vorticity at $r_c$. To correct for
nonadiabaticity and to smooth $|A(r)|^2$ near $r_c$, the
one-dimensional time independent amplitude
equation~\cite{newell:1969} is solved,
\begin{equation}
0 = \epsilon (r) A + {{\xi}_o}^2 {\cos}^2 {\theta}
\frac{\partial^2A}{\partial r^2} - g_o |A|^2A, \label{eq:amp_dim}
\end{equation}
where ${{\xi}_o}^2 = 0.148$,
$g_o=(0.6995-0.0047\sigma^{-1}+0.0083\sigma^{-2})$ and
$\epsilon(r)$ is determined by
\begin{equation}
 \epsilon(r) = \left\{ \begin{array}{ll}
   \epsilon_o, & \mbox{$r < r_0$} \\
   \epsilon_o (h^3-h_c^3)/(1-h_c^3), & \mbox{$r \ge
   r_0$}
\end{array}\right.
\label{eq:eps}
\end{equation}
where $h_c=h(r_c)$. Equation~(\ref{eq:amp_dim}) is solved
numerically using the boundary conditions $\partial_r A=0$ at $r =
0$, and $A=0$ at $r = r_1$.

\begin{figure}[tbh]
\begin{center}
\includegraphics[width=2.0in]{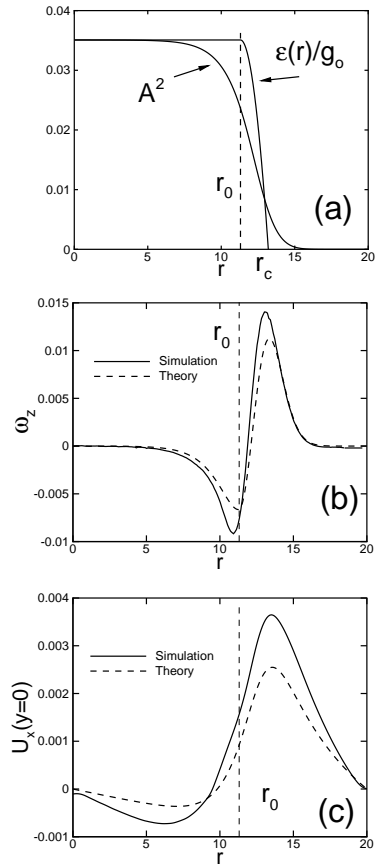}
\end{center}
\caption{Panel~(a) shows the solution of Eq.~(\ref{eq:amp_dim})
plotted as $A^2(r)$, shown for comparison is $\epsilon(r)/g_o$.
Panel~(b) compares the vertical vorticity found analytically from
Eq.~(\ref{eq:amp_dim}) with an angular average, weighted by $\sin
2 \theta$, of the vertical vorticity from simulation. Panel~(c)
compares the mean flow found analytically from
Eq.~(\ref{eq:vorticity_potential}) with the mean flow from
simulation. Parameters are $r_0=11.31$, $r_c=13.20$, $r_1=20.0$,
$\delta_r=0.036$, $\sigma=0.87$ and $\epsilon_o = 0.025$.}
\label{fig:amp_vort_mf_exp1750}
\end{figure}

To compare these analytical results with simulation we have chosen
to investigate a large-aspect-ratio cylinder with a gradual radial
ramp, defined by Eq.~(\ref{eq:platesep}), given by the parameters:
$r_0=11.31$, $r_1=20.0$, $\delta_r=0.036$, and $\sigma=0.87$. For
small $\epsilon$ the amplitude $A^2(r)$ is unable to adiabatically
follow the ramp, this nonadiabaticity results in a deviation from
$\epsilon(r)/g_o$ as shown in Fig.~\ref{fig:amp_vort_mf_exp1750}a.
However, as $\epsilon_0$ increases the amplitude $A^2(r)$ follows
$\epsilon(r)/g_o$ adiabatically almost over the entire ramp except
for a small kink at $r_c$. The structure of $\omega_z$ depends
upon this adiabaticity and is shown in
Fig.~\ref{fig:amp_vort_mf_exp1750}b where we have used the
solution to Eq.~(\ref{eq:amp_dim}) at $\theta=\pi/4$ in
Eq.~(\ref{eq:wz}). This is not strictly correct since the
non-adiabaticity of the amplitude is $\theta$ dependent which will
induce higher angular modes of the vorticity not given by
Eq.~(\ref{eq:wz}). However, the calculation should give a good
approximation to the main $\sin 2 \theta$ component of the
vorticity. It is evident from Fig.~\ref{fig:amp_vort_mf_exp1750}b
that the vertical vorticity, calculated from the simulation
results as an angular average weighted by $\sin 2 \theta$ has an
octupole angular dependence (octupole in the sense of an inner and
outer quadrupole) and is well approximated by theory without any
adjustable parameters.

The mean flow generated by these vorticity distributions is
determined by solving Eq.~(\ref{eq:wz_general}) with the boundary
condition $\zeta(r_1)=0$. The vorticity potential is related to
the mean flow in polar coordinates by $(U_r,U_\theta)=(r^{-1}
\partial_\theta \zeta, -\partial_r \zeta)$. The vorticity
potential is expanded radially in second order Bessel functions
while maintaining the $\sin 2 \theta$ angular dependence. Of
particular interest is the mean flow perpendicular to the
convection rolls, $U_r(\theta=0)$ or equivalently $U_x(y=0)$,
which is shown in Fig.~\ref{fig:amp_vort_mf_exp1750}c. Again the
simulation results compare well with theory even in the absence of
adjustable parameters.

\begin{figure}[tbh]
\begin{center}
\includegraphics[width=2.5in]{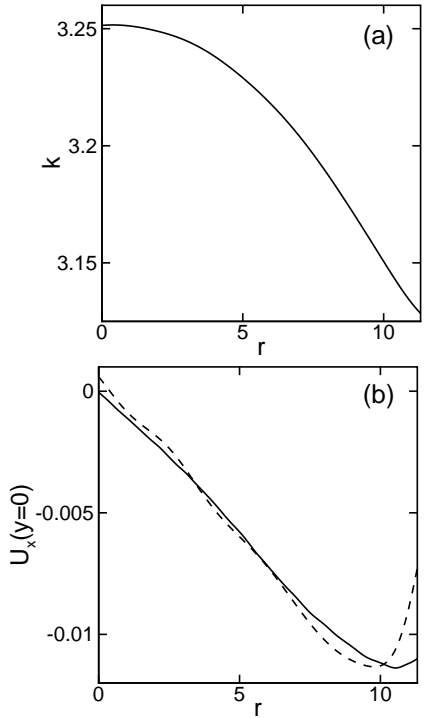}
\end{center}
\caption{Panel~(a), the variation in the local wavenumber along
the positive x-axis, or equivalently $k(r)$ at $\theta=0$.
Panel~(b), a comparison of the mean flow from simulation (solid
line) with the predicted value (dashed line) calculated from
Eq.~(\ref{eq:phase}) using the wavenumber variation from
panel~(a). Simulation parameters, $r_0=11.31$, $r_1=20$,
$\delta_r=0.036$, $\sigma=0.87$ and $\epsilon=0.171$.}
\label{fig:wn_mf_compare2000}
\end{figure}

To make the connection between mean flow and wavenumber
quantitative it is noted that the wavenumber variation resulting
from a mean flow across a field of x-rolls can be determined from
the one-dimensional phase equation,
\begin{equation} \label{eq:phase}
U \partial_x \phi = D_\parallel \partial_{xx}\phi
\end{equation}
where the wavenumber is the gradient of the phase,
$k=\partial_x\phi$, $D_\parallel=\xi_o^2 \tau_o^{-1}$, and
$\tau_o^{-1}=19.65 \sigma (\sigma+0.5117)^{-1}$~\cite{cross:1993}.
Figure~\ref{fig:wn_mf_compare2000}a illustrates the wavenumber
variation for a large-aspect-ratio simulation, $k(r)$ for $r \le
r_0$, and makes evident the roll compression, $k(r=0)>k(r_0)$.
Figure~\ref{fig:wn_mf_compare2000}b compares the mean flow
calculated from simulation to the predicted value of the mean flow
required to produce the wavenumber variation shown in
Fig.~\ref{fig:wn_mf_compare2000}a. The agreement is good and the
discrepancy near $r_0$, which is contained within one roll
wavelength from where the ramp begins, is expected because the
influence of the ramp was not included in Eq.~(\ref{eq:phase}).
This illustrates quantitatively that is in indeed the mean flow
that compresses the rolls in the bulk of the domain.

Finally, to better understand the connection between mean flow and
pattern dynamics, especially that of spatiotemporal chaotic states
exhibiting both temporal chaos as well as spatial disorder, we
apply a novel numerical procedure to eliminate mean flow from the
fluid equation, Eq.~(\ref{eq:mom}), thereby evolving the dynamics
of an artificial fluid with no explicit contributions from mean
flow. In this way, we can then obtain quantitative comparisons
between the patterns generated by this artificial fluid with mean
flow quenched and by the original fluid equation.

We have applied this procedure to study spiral defect chaos (see
bottom of Fig.~\ref{fig:large_gamma}) \cite{morris:1993}. Numerous
attempts have been made to understand how a spiral defect chaos
state is formed and how it is sustained.  For example, experiments
\cite{assenheimer:1993:prl,assenheimer:1994} have found that
spirals transition to targets when the Prandtl number is
increased.  Owing to the fact that the magnitude of mean flow is
inversely proportional to the Prandtl number, c.f.
Eq.~(\ref{eq:meanflow}), it was believed that spiral defect chaos
is a low Prandtl number phenomenon for which mean flow is
essential to their dynamics. This is supported by studies of
convection models based on the generalized Swift-Hohenberg
equation \cite{xi:1993:prl,xi:1993:pre,xi:1995}, where spiral
defect chaos is not observed unless a term corresponding to mean
flow is explicitly coupled to the equation.  However, these
observations are by themselves insufficient.  For example, there
are many other effects in the fluid equations that grow towards
low Prandtl numbers, and there could be limitations in the
Swift-Hohenberg modelling.  We have applied our numerical
procedure to this case to explicitly confirm the role of mean flow
in the dynamics of spiral defect chaos.
\begin{figure}[tbh]
\begin{center}
\includegraphics[width=1.5in]{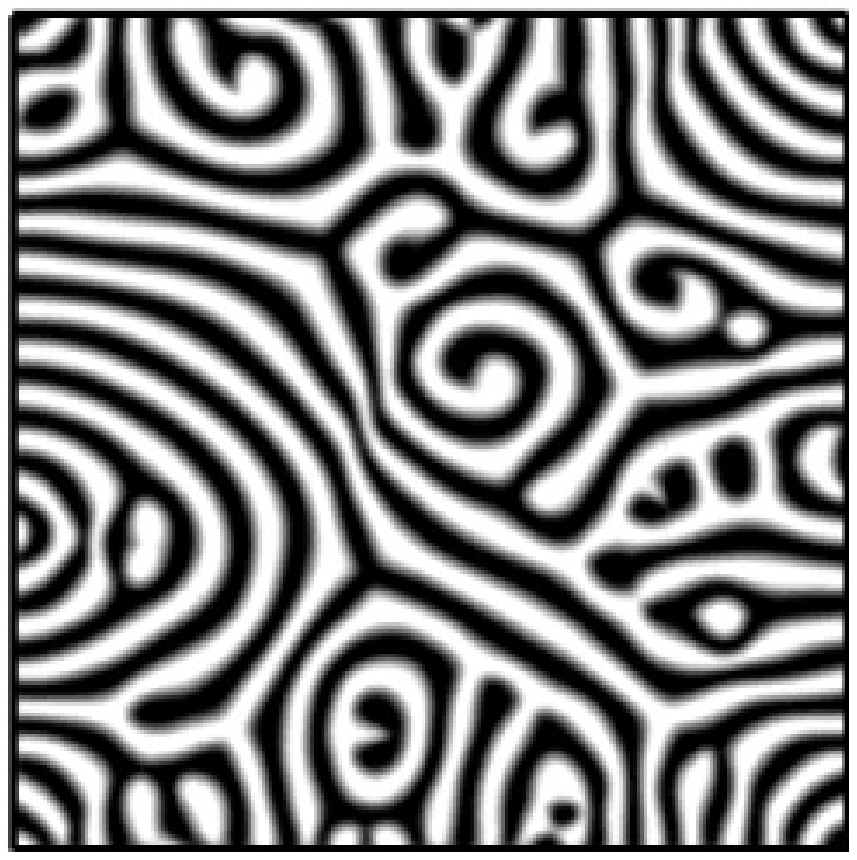}
\includegraphics[width=1.5in]{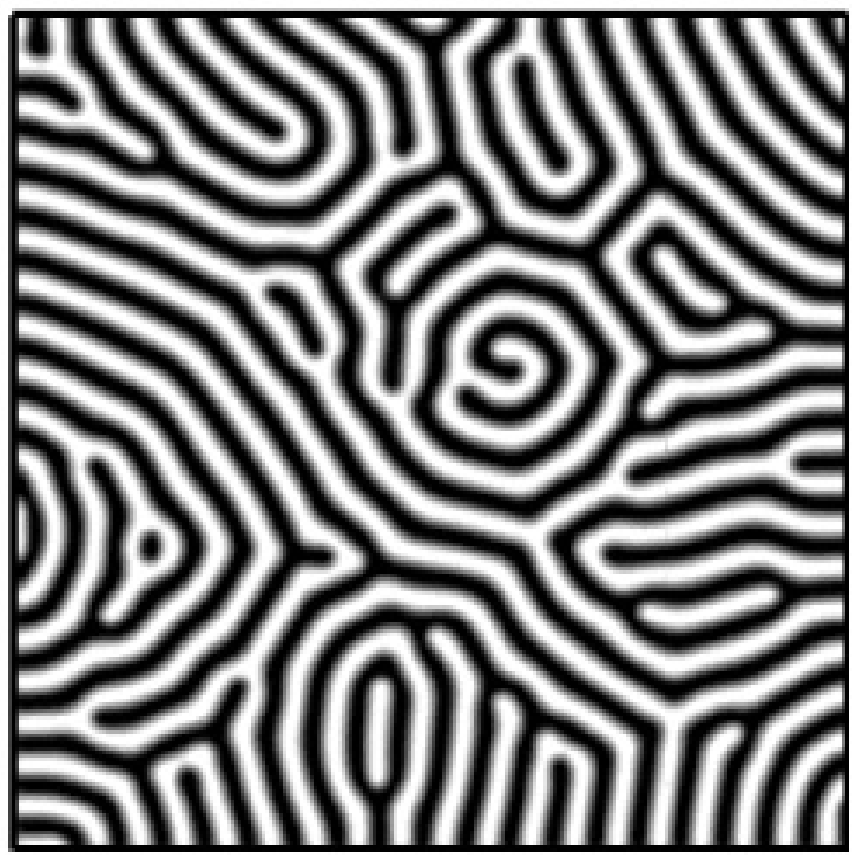}
\end{center}
\caption{Spiral defect chaos (left) and angular textures (right)
obtained by quenching mean flow.  The left panel is at
$t=152\tau_v$ and displays the pattern upon which the mean flow is
quenched, the right panel is at $t=320\tau_v$. In both cases,
$R=2950, \sigma=1.0$ and the lateral sidewalls are insulating. We
see that the spiral arms transition to angular textures when mean
flow is quenched. Also, the quenched state is stationary.}
\label{fig:quench}
\end{figure}

Recalling that we can approximate mean flow to be the
depth-averaged horizontal velocity, c.f.
Eq.~(\ref{eq:mean_flow_sim}), we can first depth-average the
horizontal components of the fluid equation, Eq.~(\ref{eq:mom}),
to obtain a dynamical equation for the mean flow $\vec{U}_s$:
\begin{eqnarray}
  \lefteqn{\sigma^{-1} \partial_t \vec{U}_s + \sigma^{-1} \int_0^1 dz
  (\vec{u}\dotprod\vec{\nabla}) \vec{u}_\perp =} \nonumber \\
 & & -\vec{\nabla}_\perp \int_0^1 dz p +
  \nabla_\perp^2 \vec{U}_s + \int_0^1 dz \partial_{zz} \vec{u}_\perp.
\end{eqnarray}
In this equation, the term $-\nabla_\perp \int_0^1 dz p$ can be
absorbed into the nonlinear Reynolds stress term via
Eq.~(\ref{eq:pressure_p0}) and so will be ignored henceforth.  The
resulting equation is then a diffusion equation in $\vec{U}_s$
with a source term $\vec{F}_s \equiv \int_0^1 dz
(\vec{u}\dotprod\nabla) \vec{u}_\perp -\sigma \int_0^1 dz
\partial_{zz} \vec{u}_\perp$.  If this source term were not
present, then $\vec{U}_s$, being the solution to a diffusion
equation, evolves to zero with an effective diffusivity $\sigma$,
the Prandtl number.  Thus, the role of $\vec{F}_s$ is to act as a
generating source for the mean flow $\vec{U}_s$.  Subtracting it
from the fluid equation, Eq.~(\ref{eq:mom}), then results in the
mean flow being eliminated.

In practice, we found that it is necessary to actually subtract
$\vec{F}_s$ multiplied by a constant to ensure that the magnitude
of mean flow becomes zero.  This can be understood in terms of the
necessity to correct for the fact that
Eq.~(\ref{eq:mean_flow_sim}) is only an approximation to the flow
field that advects the rolls given by
\begin{equation}
  \vec{U} = \int_0^1 dz g(z) \vec{u}_\perp
\end{equation}
where $g(z)$ is a weighting function depending on the full
nonlinear structure of the rolls.  This is discussed further
elsewhere \cite{chiam:2002}.

We have carried out this procedure by introducing the term
$\vec{F}_s$ to the right-hand-side of the fluid equation after a
spiral defect chaotic state becomes fully developed, typically
after about one horizontal diffusion time starting from random
thermal perturbations as initial condition.  We see that the
spirals immediately, on the order of a vertical diffusion time,
``straighten out'' to form angular chevron-like textures; see
Fig.~\ref{fig:quench}. Unlike spiral defect chaos, these angular
textures are stationary (with the exception of the slow motion of
defects such as the gliding of dislocation pairs). Thus, we have
shown that when mean flow is quenched via the subtraction of the
term $\vec{F}_s$ from the fluid equation, spiral defect chaos
ceases to exist.

We have further quantified the differences between spiral defect
chaos and the angular textures.  We mention here briefly one of
the results: by comparing the wavenumber distribution for both
sets of states, we have observed that the mean wavenumber
approaches the unique wavenumber possessed by axisymmetric
patterns asymptotically far away from the center
\cite{buell:1986:axi}.  (The axisymmetric pattern, by symmetry,
does not have mean flow components.)  We discuss this as well as
other results in a separate article \cite{chiam:2002}.
\section{Conclusion}
\label{section:conclusion} Full numerical simulations of
Rayleigh-B\'{e}nard convection in cylindrical and rectangular
shaped domains for a range of aspect ratios, $5 \lesssim \Gamma
\lesssim 60$, with experimentally realistic boundary conditions,
including rigid, finned and spatially ramped sidewalls, have been
performed. These simulations provide us with a complete knowledge
of the flow field allowing us to quantitatively address some
interesting open questions.

In this paper we have emphasized the exploration of the mean flow.
The mean flow is important in a theoretical understanding of the
pattern dynamics, yet is very difficult to measure in experiment,
making numerical simulations attractive to close this gap.

The mean flow is found to be important in small cylindrical
domains by investigating the result of imposing different sidewall
boundary conditions. Analytical results are developed for a
large-aspect-ratio cylinder with a radial ramp in plate
separation. Numerical results of the vertical vorticity and the
mean flow agree with these predictions. Furthermore, the
wavenumber behavior predicted using the mean flow in a
one-dimensional phase equation also agrees with the results of
simulation. This allows extrapolation of the analysis to larger
aspect ratios.

Lastly we utilize the control and flexibility offered by numerical
simulation to investigate a novel method of quenching numerically
the mean flow. We apply this to a spiral defect chaos state and
find that the time dependent pattern becomes time independent,
angular in nature, and that the pattern wavenumber becomes larger.

These quantitative comparisons illustrate the benefit of
performing numerical simulations for realistic geometries and
boundary conditions as a means to create quantitative links
between experiment and theory.

We are grateful to G. Ahlers for helpful discussions. This
research was supported by the U.S.~Department of Energy, Grant
DE-FT02-98ER14892, and the Mathematical, Information, and
Computational Sciences Division subprogram of the Office of
Advanced Scientific Computing Research, U.S.~Department of Energy,
under Contract W-31-109-Eng-38. We also acknowledge the Caltech
Center for Advanced Computing Research and the North Carolina
Supercomputing Center.

\end{document}